\documentclass{PoS}
\usepackage{graphicx}

\makeatletter
\newbox\slashbox \setbox\slashbox=\hbox{\large$/$}
\def\pslash#1{\setbox\@tempboxa=\hbox{$#1$}
  \@tempdima=0.5\wd\slashbox \advance\@tempdima 0.5\wd\@tempboxa
  \copy\slashbox \kern-\@tempdima \box\@tempboxa}
\def\slash{\protect\pslash}
\makeatother

\newcommand{\mat}{\left ( \begin{array}{cc}}
\newcommand{\emat}{\end{array} \right )}
\newcommand{\be}{\begin{eqnarray}}
\newcommand{\ee}{\end{eqnarray}}
\newcommand{\ba}{\begin{array}}
\newcommand{\ea}{\end{array}}

\def\SB{S$\chi$SB}

\def\beq{\begin{equation}}
\def\eeq{\end{equation}}

\title{Chiral phase transition as an Anderson transition
 in the Instanton Liquid Model for QCD}
\ShortTitle{Chiral phase transition as an Anderson transition in the ILM}
\author{\speaker{Antonio M. Garc\'{\i}a-Garc\'{\i}a}\\

Physics Department, Princeton University, Princeton,
New Jersey 08544, USA\\
E-mail:\email{ag3@princeton.edu}}

\author{James C. Osborn \\

Physics Department \& Center for Computational Science,
 Boston University, Boston, MA 02215, USA\\
E-mail:\email{josborn@physics.bu.edu} }

\abstract{
We study the spectrum and eigenmodes of the
 QCD Dirac operator in a gauge background given by an Instanton Liquid Model
 (ILM) at temperatures around the chiral phase transition.
For two massless quark flavors
 we observe that at the chiral phase transition, both the 
 low lying eigenmodes and the spectrum of the QCD Dirac operator 
 resemble those of a disordered conductor at the 
 Anderson transition.
In particular we present results showing that the eigenvectors are
 multifractal and the spectral
 correlations are well described by critical statistics. 
We also find an additional temperature dependent mobility 
 edge located in the
 bulk of the spectrum with properties quantitatively similar to those
 of a 3D disordered conductor at the Anderson transition. 
This feature is
 present in both quenched and unquenched simulations.
Finally we argue that our findings are not in principle restricted to the ILM
 approximation and may also be found in lattice simulations.}
\FullConference{XXIIIrd International Symposium on Lattice Field Theory\\

                 25-30 July 2005\\

                 Trinity College, Dublin, Ireland}
 \PoS{PoS(LAT2005)265} 
\begin{document}

\section{Introduction} 

The spontaneous chiral symmetry breaking (\SB) in 
 QCD and its eventual restoration at finite temperature is one of the 
 most intensively investigated topics in hadronic physics.
QCD models whose gauge configurations
are given by an interacting liquid of instantons \cite{shuryak}
provide an adequate theoretical framework to understand \SB.
Thus, based on the semiclassical picture of a QCD vacuum 
 dominated by instantons, it has been suggested  
\cite{chisbinst}
  that the
\SB{} in QCD and  the phenomenon of conductivity  
 may have similar physical origins.
Conductivity in a disordered sample is produced by electrons that although
initially bound to an impurity may become delocalized by orbital overlapping
 with nearby impurities. 
Similarly, in the QCD vacuum, the zero modes of the Dirac operator though 
initially bound to an instanton may get delocalized due to  
the strong overlap with other instantons. 
As a consequence, the chiral condensate becomes nonzero and
chiral symmetry is spontaneously broken. 
In this paper we provide evidence that
 these analogies can be extended to describe the 
 chiral phase transition in QCD at finite temperature. 
Specifically, in the context of an Instanton Liquid Model (ILM),
we show that around the temperature that the chiral 
phase transition occurs the low lying eigenmodes
 of the QCD Dirac operator undergo an Anderson transition (AT), 
namely,
 a localization-delocalization transition, characterized by 
 multifractal eigenstates and critical statistics \cite{anderson}.

For the sake of clarity we
briefly summarize the 
properties of a disordered system at an AT.
In three and higher dimensions there exists
 a mobility edge separating localized from delocalized states.  
In the delocalized region, eigenfunctions are extended through the sample and
 the level statistics are described by random matrix theory (RMT).  
In the opposite limit, eigenfunctions
 are exponentially localized and the spectral correlations are 
described 
by Poisson statistics.       
Around the mobility edge, the region where the AT occurs,
 the eigenstates are multifractal, meaning that moments of the eigenfunctions
  exhibit anomalous scaling with respect 
  to the sample size \cite{aoki}. 
Typical features of the
 level statistics at the AT, dubbed critical statistics \cite{kravtsov97},
include scale invariant spectrum \cite{sko},
 level repulsion and asymptotically linear number 
 variance.
Thus the level spacing distribution goes to zero as the spacing goes to zero
 as in a metal but long range correlators such as the number variance
 $\Sigma^2(\ell)=\langle (N_\ell -\langle N_\ell \rangle)^2 \rangle
  \sim \chi \ell$
 for $\ell \gg 1$ ($N_\ell$ is the number of eigenvalues in an interval of
length $\ell$)
 are asymptotically linear, as for an insulator ($\chi = 1$), but with a
 slope $\chi < 1$ ($\sim 0.27$ for a 3D Anderson model \cite{chiat}).
Below we show under what circumstances the above features are also found
in the ILM at finite temperature. 

\begin{figure}[ht]
  \hfill
  \begin{minipage}[t]{.45\textwidth}
    \begin{center}  
      \includegraphics[width=\columnwidth]{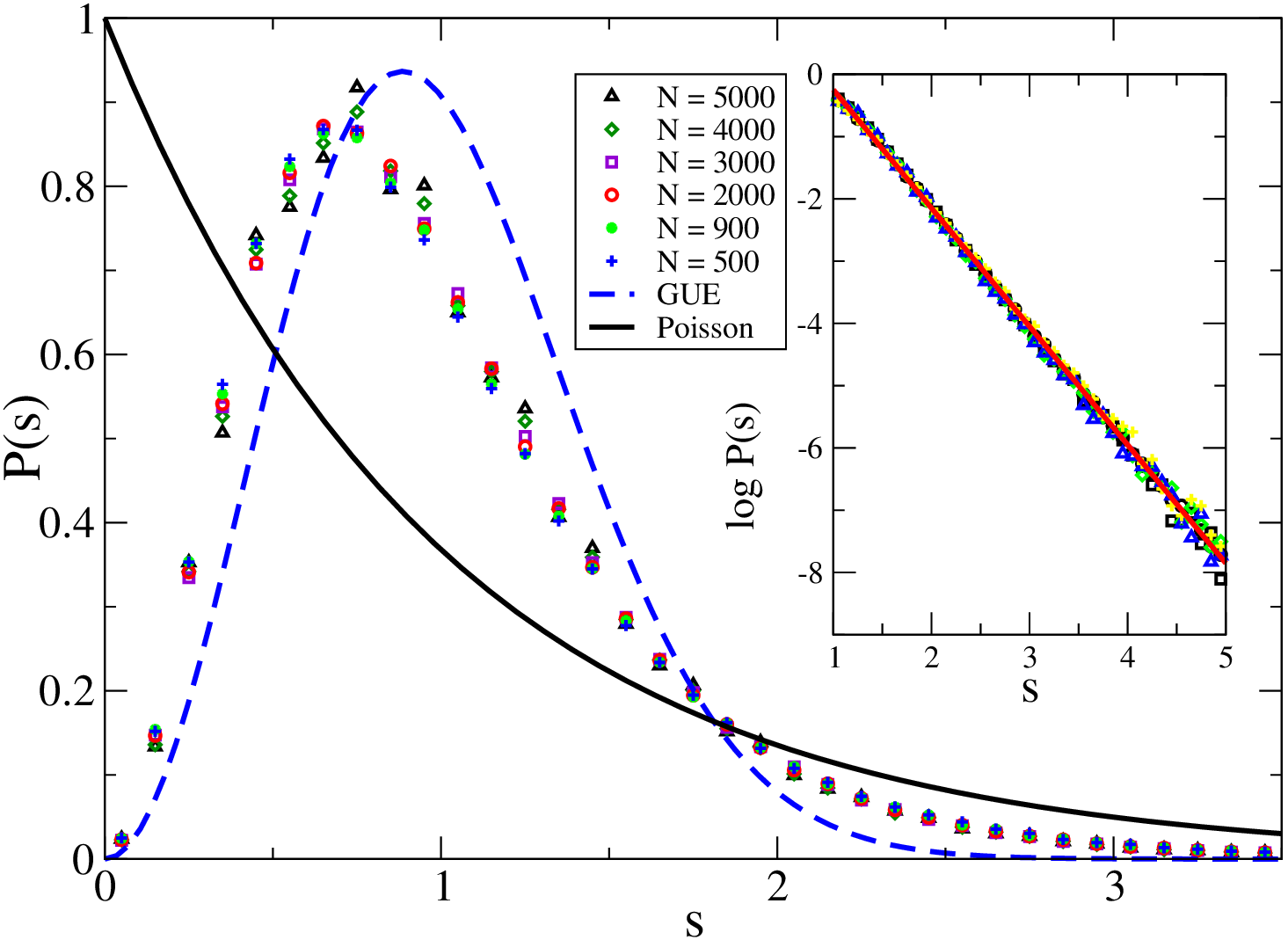}
      \caption{Level spacing distribution $P(s)$ in the bulk for the quenched
               ILM at $T=200$ MeV for different volumes.
               The inset shows $\log P(s)$ for the tail of the data.
               The best fit (solid line) corresponds to a slope $-1.64$.}
      \label{nf0ps1}
    \end{center}
  \end{minipage}
  \hfill
  \begin{minipage}[t]{.45\textwidth}
    \begin{center}  
      \includegraphics[width=\columnwidth]{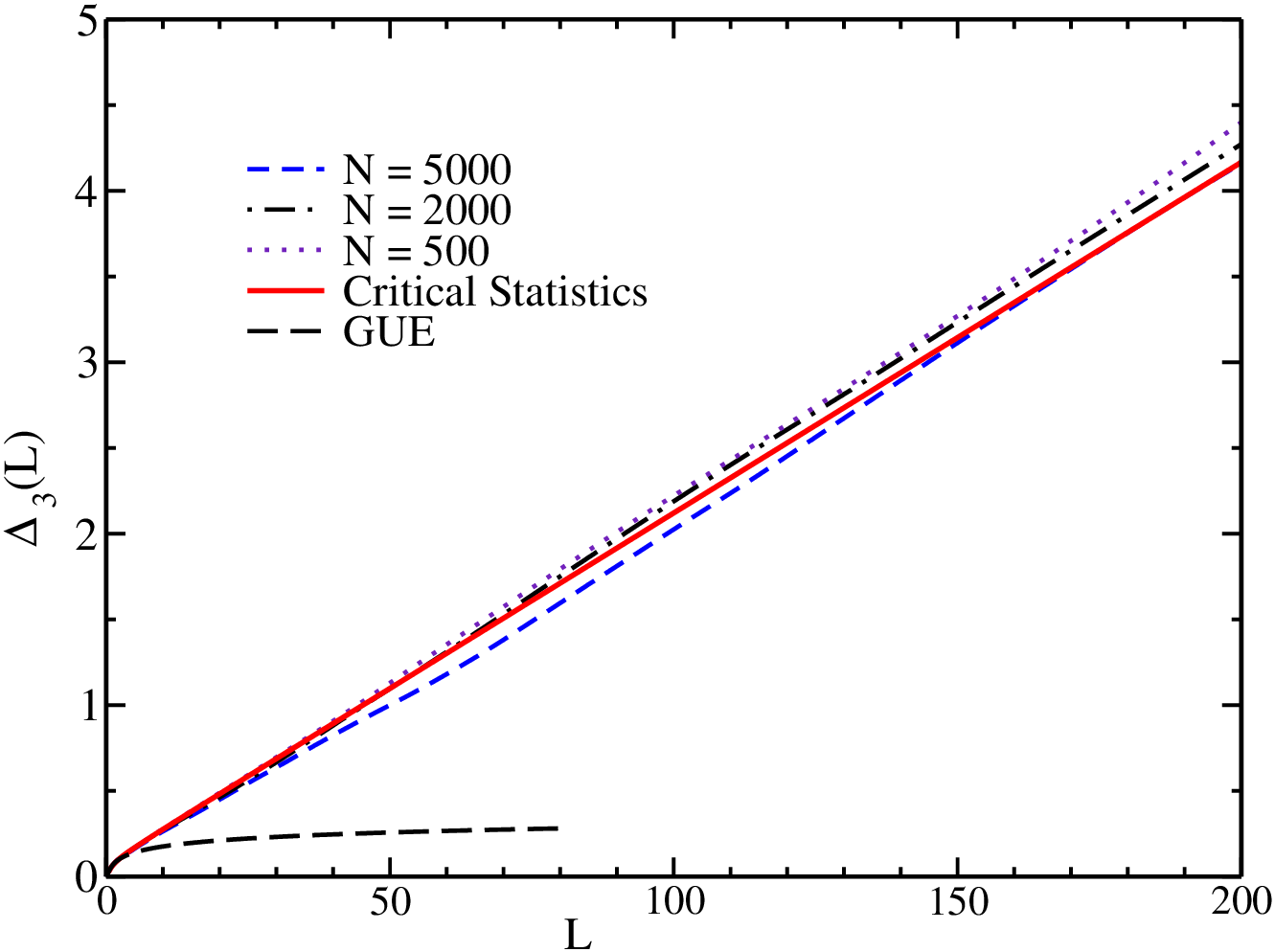}
      \caption{Spectral rigidity 
in the bulk for the quenched ILM at $T=200$ MeV
 for different volumes.
 The result has very little size dependence and agrees well with
 the prediction of critical statistics (solid line).}
      \label{nf0d3}
    \end{center}
  \end{minipage}
  \hfill
\end{figure}

\section{The instanton liquid partition function: Technical details}

Our starting point
 is the Euclidean QCD partition function,
\be
Z_{\rm inst} = \int D\Omega \, {\det}^{N_f} (\slash{D} + m) e^{-S_{\rm YM}},
\label{zinst}
\ee
where the integral is over the collective coordinates of the instantons,
 the Dirac operator is denoted by $\slash{D}$ and
 the Yang-Mills action is denoted by $S_{\rm YM}$.  
For each instanton we have $12$ collective coordinates (for three colors).
The fermion determinant is evaluated in the space
of the fermionic zero modes of the instantons (for further details we refer to 
 the review \cite{SS97}).
For simplicity we have kept the density fixed at $N/V=1\,\mathrm{fm}^{-4}$
as the temperature is increased.
This is justified since it was already observed that the density of instantons
at the transition is still sizable, around $0.6\,\mathrm{fm}^{-4}$,
and the transition is not expected to be driven by the decrease in density.
We are therefore using this as a qualitative model for the QCD
chiral phase transition even though the transition temperature will differ
from the lattice prediction.

We report results for $N_f = 0$ (quenched) and
 $N_f=2,~m_u=m_d=0$ (unquenched). The effect of a nonzero mass 
 or a strange quark will be discussed elsewhere \cite{aj2}.
The partition function (\ref{zinst}) is evaluated by means of a
Metropolis algorithm. In all simulations we performed between 500 and
1000 equilibration updates followed by 2000 to 10000 measurement updates.
In the quenched case results are presented for ensembles of up to
 $N=5000$ instantons and anti-instantons.
In the unquenched case, the simulations are slower and we have only 
investigated ensemble of up to $N=504$.
All the spectral correlators are calculated from the unfolded spectrum.
This procedure scales the eigenvalues so that the average density on the scale
of several level spacings is one.

\begin{figure}[t]
\centering
\includegraphics[width=0.6\columnwidth,clip]{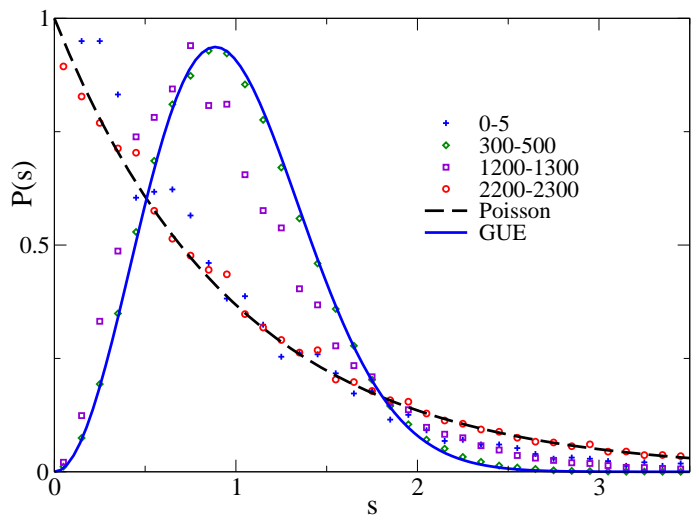}
\caption{Level spacing distribution $P(s)$ for the quenched ILM
 at $N=5000$ and $T=200$ MeV for different spectral windows.
 The ranges indicate the eigenvalue numbers plotted with 0 being the smallest
 eigenvalue.}
\label{nf0ps}
\end{figure}

\section{Localization in the Instanton Liquid Model: The quenched case}

Here we examine the properties of eigenstates in the center of the spectrum
 (bulk).  Since only the smallest eigenmodes are responsible for \SB,
the mobility edge seen here is not related to the chiral transition
(strictly speaking there is not a chiral transition in the quenched case
 anyway).
We instead study it because it is a very clear example of a mobility edge
 in the ILM and we find it has similar features to the mobility 
edge 
near origin, but since it 
affects a larger range of eigenmodes we can get better
 statistics.
By using the finite scaling method introduced in \cite{sko} we have observed a
 mobility edge separating localized from delocalized eigenstates
 in the range $T \sim$ 150 -- 250 MeV. As the temperature
decreases the location of the mobility edge moves to the end of the
spectrum. For $T < 170$ MeV the results are less reliable since the
mobility edge is located almost in the end of the spectrum where
truncation effects are larger. 
  Following the literature in
disordered systems we have investigated the temperature $T \sim 200$ MeV
such that the mobility edge is located around the center of the
spectrum. The level statistics in this spectral region 
have the signatures of an AT including a
spectrum that is scale invariant to high degree. As observed in
 figures \ref{nf0ps1} and \ref{nf0d3} both short range statistics such as
the level spacing distribution, $P(s)$, and long range 
 correlators such as the spectral rigidity
  $\Delta_{3}(\ell)=\frac{2}{\ell^4}\int_{0}^{\ell}(\ell^3-2\ell^2x+x^3)\Sigma^{2}(x)dx$
do not depend on the system size
 for volumes (number of instantons) ranging from $N=500$ to $N=5000$.
 Level repulsion is still present as for a 
 metal but the spectral rigidity (see figure \ref{nf0d3}) is asymptotically
 linear with a slope that corresponds to $\chi \sim
 0.29 \pm 0.02$ in fair agreement with the value for a 3D disordered system at
 an AT of $\chi \sim 0.27\pm 0.02$ \cite{chiat}. Furthermore the 
 exact form of the spectral rigidity follows closely the prediction 
 of critical statistics, equation (31) in \cite{ant4} with $h$,
 a free parameter, set to a value of $0.62$.
 As observed in figure \ref{nf0ps}
the level statistics outside the critical region are
 RMT like in the region between the origin and the mobility edge
 and Poisson like between the mobility edge and the end of the spectrum. 
 
We now present the eigenvector properties. A
signature of an AT is the multifractality of the
eigenstates as measured by the moments
 $P_q=\int d^dr |\psi({\bf r})|^{2q}\propto L^{-D_q(q-1)}$,
 where $D_q$ is a set of different exponents describing the
 transition and $L \propto N^{1/3}$ is the spatial system size.
In order not to consider eigenstates
outside the critical region we have taken only $10\%$ of the
eigenvectors around the center of the spectrum for $T=200$ MeV. 
By fitting $P_2$ (also known as the inverse participation ratio or IPR)
for different volumes, 
 we have obtained a value of $D_2 \sim 1.5 \pm 0.1$ in 
 agreement with that of a 3D disordered system at the AT \cite{mirlin}.

Our numerical analysis shows that at $T=200$ MeV there is mobility edge
 around the central part of the spectrum in the quenched ILM.
The level statistics have all the signatures of an Anderson transition:
 scale invariance, level repulsion and sub-Poisson spectral rigidity;
 and the eigenstates appear to be multifractal.
We have also observed similar critical features in the
 region close to the origin at temperatures around 110--140 MeV.
However the analysis is complicated by the accumulation of very small
eigenvalues that are present in the quenched ILM.
More details of this region will be given elsewhere \cite{aj2}.
Instead we present results near the origin for two massless flavors 
where the accumulation of very small eigenmodes is suppressed and
the analysis is cleaner.

\begin{figure}[t]
  \hfill
  \begin{minipage}[t]{.45\textwidth}
    \begin{center}  
\includegraphics[width=\columnwidth,clip,angle=0]{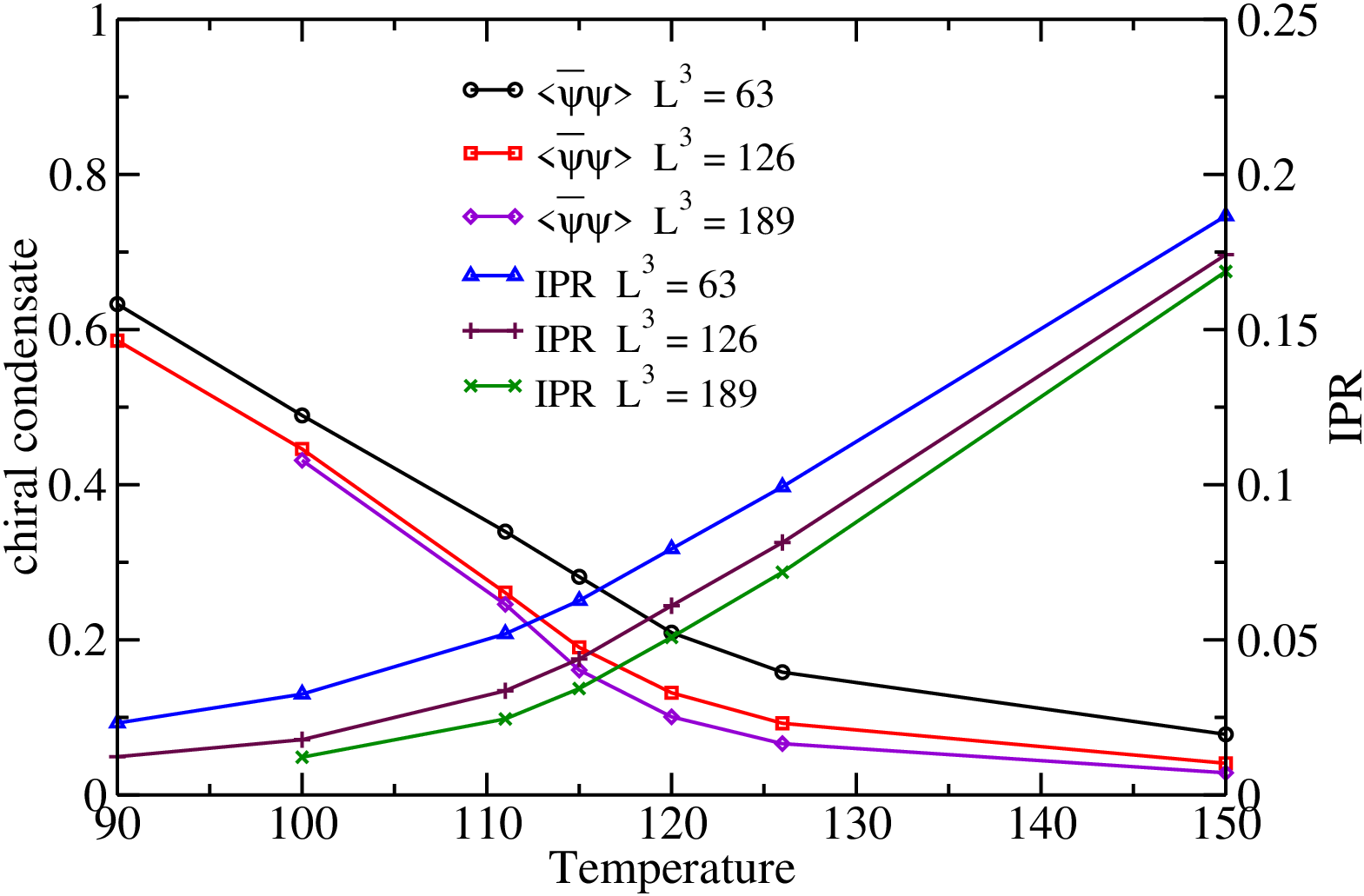}
\caption{Chiral condensate and inverse participation ratio (IPR) of the lowest
 eigenmode for the ILM with two massless quarks.}
\label{cciprnf2m0}
    \end{center}
  \end{minipage}
  \hfill
  \begin{minipage}[t]{.45\textwidth}
    \begin{center}  
\includegraphics[width=\columnwidth,clip,angle=0]{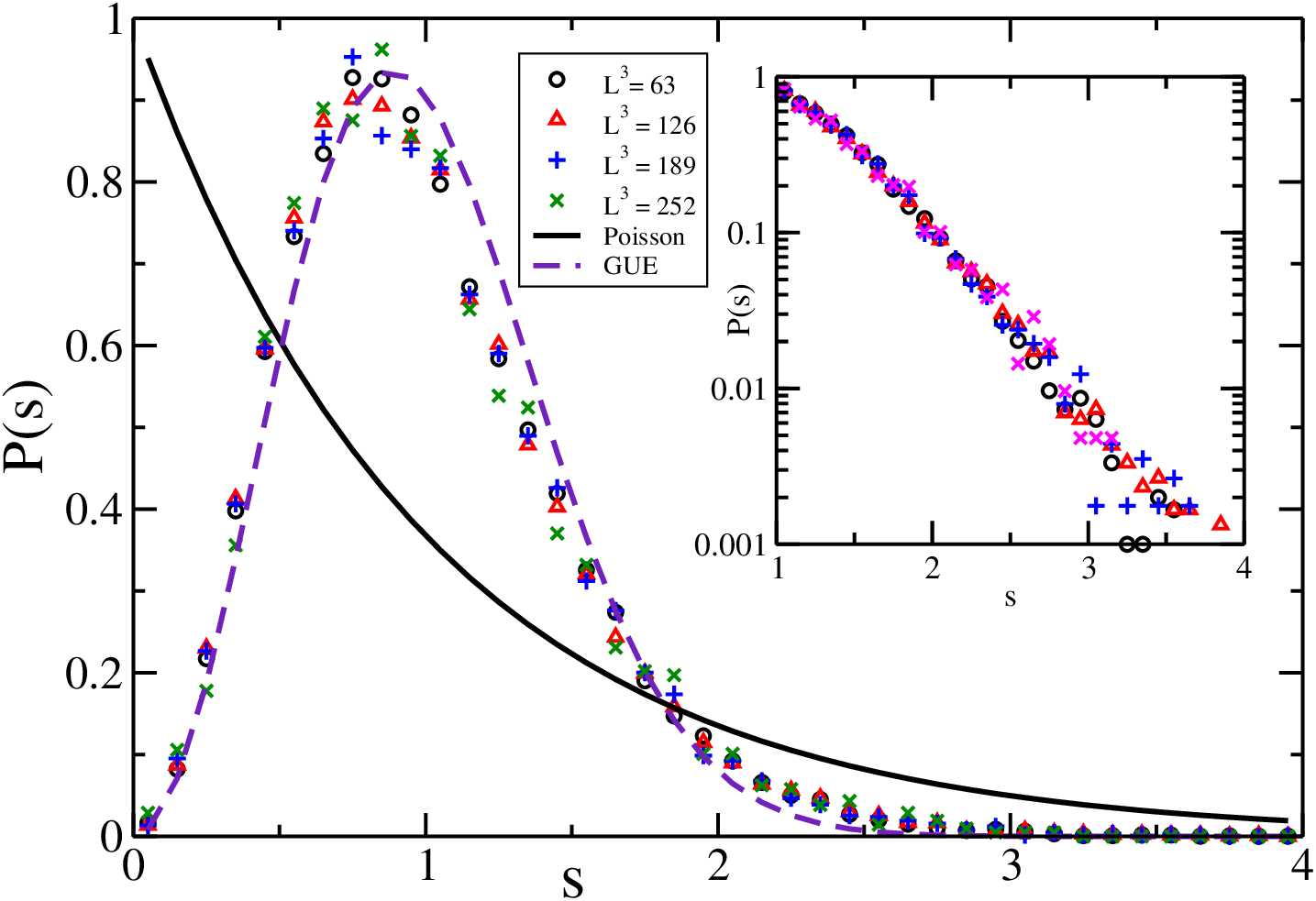}
\caption{Level spacing distribution $P(s)$ at the origin for the unquenched ILM
 at $T=115$ MeV for different system sizes.}
\label{psnf2m0}
    \end{center}
  \end{minipage}
  \hfill
\end{figure}

\section{Chiral restoration and Anderson transition: The unquenched case}

We now study eigenvalues and eigenvectors of the QCD
Dirac operator in an ensemble with two massless quark flavors.
In this case the chiral condensate associated with \SB{} is finite.
Around the critical temperature, the condensate approaches zero signaling
 the chiral phase transition.
Since we have two massless flavors we expect to find a
 second order chiral phase transition.

We focus our attention on the spectral 
 region close to origin since our aim is to relate the chiral phase transition 
 with a transition to localization (AT) of the QCD Dirac operator eigenmodes. 
By looking at the chiral condensate versus temperature for a range
of system sizes (figure \ref{cciprnf2m0}) we see that the condensate
approaches zero around temperatures of 115 -- 120 MeV in a manner consistent
with a second order phase transition.
At the same time the condensate is falling we see the IPR of the lowest
 eigenmode begin to rise signaling a transition to localization. 
This finding strongly suggests that both phenomenon are
 intimately related.

Next we examine the nature of the localization transition.
First we locate, by using the finite size scaling method \cite{sko},
 the AT for the lowest lying eigenmodes at $T \sim$ 115 -- 120 MeV.
At this temperature we see that eigenvalue statistics such as the
level spacing distribution (figure \ref{psnf2m0}) are nearly scale
invariant, there is level repulsion but the tail of $P(s)$ is exponential as at the AT.
We also looked at the scaling of $P_q$ in the lowest eigenmode
 at $T=120$ MeV for system sizes ranging from $L^3=$63--252.
This yielded a set of fractal dimensions
$D_2=1.3\pm0.2$, $D_3=0.9\pm0.2$, $D_4=0.7\pm 0.3$ and $D_5=0.7\pm0.3$
showing that the eigenstates are multifractal.
Although higher volumes would be desirable to 
fully confirm the multifractal scaling of the eigenstates, it is
encouraging that the IPR of comparatively similar temperature
 ($T \sim 100$ Mev) is comparable to the one for a fully metallic sample.
We also mention that we observed a
 mobility edge on the bulk of the spectrum at a slightly higher
 temperature with properties similar to the quenched case.

At this point a natural question to address is  
 whether the results here reported concerning the Anderson transition 
 in the ILM may also be present in more 
 realistic models of the strong interactions.
Although no conclusive answer can be given to this question
 we would like to point out the 
 reasons why we have found an Anderson transition in the ILM.
As is known, unlike the zero temperature case where the decay is power-law, 
 the fermionic zero modes in the field of an instanton at nonzero temperature
 (usually called a caloron)
 have an exponential tail $e^{-rT}$ in the spacial directions and are
 oscillatory in the time direction \cite{SS97}.
This suggests that the overlap among different zero modes is essentially
 restricted to nearest neighbors in the spacial directions.
However, in the time direction different zero modes strongly overlap
 due to the oscillatory character of the eigenmodes.
This situation strongly resembles a 4D disordered conductor 
 in the tight binding approximation (only nearest neighbor hopping) 
 with one dimension (time) much smaller than the rest so the system can be
 considered effectively three dimensional.
It is well established that such a system may undergo an AT depending on the disorder
 strength.
In our case the disorder role is played by temperature since the
wavefunction decays as $\sim e^{-rT}$.  From this discussion it is natural
 to find an Anderson transition in the
  ILM for a particular value of the temperature.
The principal ingredient to reach the AT is an exponential decay
 of the eigenmodes explicitly depending on the temperature together with the 
 possibility to tune the effective range of the exponential in order to 
 reach the transition region.
Any theory with these features very likely 
 will undergo an AT for some value of the parameters.
It is therefore quite possible that the QCD vacuum has similar properties.
Even if the objects responsible for localization are not the classical
 instantons (see the topology section of these proceedings for other options)
 this scenario could still be realized
 if the interactions behave in a similar way.
We refer to \cite{aj2} for a more detailed 
 account of this interesting issue.    

In summary we have found that in the unquenched ILM the chiral
restoration transition occurs around the same temperature as the localization
transition close to the origin. From a physical point of view this is
not surprising since the appearance of the condensate is deeply linked
with the delocalization of the zero modes in the instanton vacuum due to
long range instanton induced interactions.  
It will be very interesting to repeat these same studies in lattice
simulations where the chiral transition also coincides with a
deconfinement transition and additionally to see if the detailed structure
of the topological objects
 affects the nature of the localization transition.


\begin{thebibliography}{9}

\bibitem{shuryak} E. Shuryak, \emph{Nucl. Phys.} {\bf B203} (1982) 93,116,140.

\bibitem{chisbinst}
 D. Diakonov and P. Petrov, \emph{Nucl. Phys.} {\bf B272}
 (1986) 457; \emph{Phys. Lett.} {\bf B147} (1984) 351; hep-ph/9602375;
 J.C. Osborn and J.J.M. Verbaarschot, \emph{Phys. Rev. Lett.}
 {\bf 81} (1998) 268; \emph{Nucl. Phys.} {\bf B525} (1998) 738;
 R.A. Janik, M.A. Nowak, G. Papp, and I. Zahed,
 \emph{Phys. Rev. Lett.} {\bf 81} (1998) 264.

\bibitem{anderson} P.W. Anderson, \emph{Phys. Rev.} {\bf 109} (1958) 1492.

\bibitem{aoki} H. Aoki, \emph{J. Phys.} {\bf 16C}, (1983) L205;
 F. Wegner, \emph{Z. Phys. B} {\bf 36} (1980) 209.

\bibitem{kravtsov97} V.E. Kravtsov and K.A. Muttalib, \emph{Phys. Rev. Lett.}
 {\bf 79} (1997) 1913.

\bibitem{sko}B.I. Shklovskii, et.al.,
\emph{Phys. Rev. B} {\bf 47} (1993) 11487.

\bibitem{chiat} D. Braun, G. Montambaux and M. Pascaud,
 \emph{Phys. Rev. Lett.} {\bf 81} (1998) 1062.

\bibitem{SS97} T. Sch\"afer and E. Shuryak, \emph{Rev. Mod. Phys.}
 {\bf 70} (1998) 323.

\bibitem{aj2} A.M. Garcia-Garcia and J.C. Osborn, in preparation.

\bibitem{ant4} A.M. Garcia-Garcia and J.J.M. Verbaarschot,
 \emph{Phys. Rev. E} {\bf 67} (2003) 046104.

\bibitem{mirlin} T. Ohtsuki and T. Kawarabayashi,
 \emph{J. Phys. Soc. Jpn.} {\bf 66} (1997) 314. 

\end{thebibliography}
\end{document}